\documentclass[aps, prd, reprint, nofootinbib, nobibnotes]{revtex4-1}
\usepackage{amsmath, amssymb, graphicx, color, mathrsfs, array, bbm, hyperref}
\graphicspath{{./}{./figures/}}
\newcommand{\beq}{\begin{equation}}
\newcommand{\eeq}{\end{equation}}

\newcommand{\eq}[1]{Eq.~(\ref{#1})}
\newcommand{\fig}[1]{Fig.~\ref{#1}}
\newcommand{\sect}[1]{Sec.~\ref{#1}}

\newcommand{\rank}{\text{rank}\,}

\newcommand{\sgn}[1]{\text{sgn}\left( #1 \right)}

\newcommand{\abs}[1]{\left| #1 \right|}
\newcommand{\norm}[1]{\left\Vert #1 \right\Vert}

\newcommand{\vecenv}[2]{\left( \begin{array}{c} #1 \\ #2 \end{array} \right)}
\newcommand{\diag}[1]{\text{diag}\left\{ #1 \right\}}

\newcommand{\wt}[1]{\widetilde{#1}}

\newcommand{\cmplx}{\mathbb{C}}
\newcommand{\intg}{\mathbb{Z}}
\newcommand{\real}{\mathbb{R}}
\newcommand{\torus}{\mathbb{T}}

\newcommand{\id}{\mathbbm{1}}
\newcommand{\nullv}{\mathbf{0}}

\newcommand{\SO}{\mathrm{SO}}

\newcommand{\SL}{\mathrm{SL}}

\newcommand{\hlt}{\mathcal{H}}

\newcommand{\green}{\mathcal{G}}

\newcommand{\ve}{\varepsilon}

\newcommand{\ket}[1]{| #1 \rangle}

\renewcommand{\c}{c^{\phantom{\dagger}} }

\newcommand{\vb}{\mathbf{b}}
\newcommand{\vc}{\mathbf{c}}

\newcommand{\vk}{\mathbf{k}}

\newcommand{\vn}{\mathbf{n}}

\newcommand{\vv}{\mathbf{v}}
\newcommand{\vw}{\mathbf{w}}

\newcommand{\vcd}{\mathbf{c}^\dagger}

\newcommand{\bet}{{\boldsymbol{\eta}}}

\renewcommand{\dim}[1]{\mathrm{dim}\left( #1 \right)}

\newcommand{\dg}{{\dagger}}
\newcommand{\pdg}{{\phantom{\dagger}}}
\newcommand{\viz}{\emph{viz}}
\newcommand{\hc}{\text{H.c}}

\newcommand{\rot}{\mathcal{R}}
\newcommand{\param}{\nu}

\newcommand{\hlts}{\eta}
\newcommand{\hltsq}{\Gamma}

\newcommand{\jac}{\mathbb{J}}

\newcommand{\refsym}{b}
\newcommand{\refvec}{\vb}

\begin{document}
\title{Fermi arc reconstruction at junctions between Weyl semimetals}

\author{Vatsal Dwivedi}
\email{vdwivedi@thp.uni-koeln.de}
\affiliation{Institute for Theoretical Physics, University of Cologne, 50937 Cologne, Germany}

\begin{abstract}
  We analyze junctions between noninteracting fermionic topological phases. A closed-form condition for the existence of localized modes at the interface is derived using the transfer matrix approach. These analytical conditions as well as numerical exact diagonalization are used to study interfaces between Weyl semimetals. We observe a \emph{Fermi arc reconstruction} at the interface, leading to closed curves of zero energy modes in the interface Brillouin zone. These are stable even in certain cases where the two Weyl semimetals differ only in their Fermi arc connectivities. 
\end{abstract}

\maketitle


\section{Introduction}
The realization of topological aspects of the band theory of solids has opened new avenues in the study of insulators. A particularly dramatic manifestation of the topology is the presence of \emph{boundary modes}, i.e, single particle states exponentially localized at surface/edge of a topologically nontrivial insulator with energies lying in the bulk gap. These modes are protected ``by the topology'' and cannot be removed by local perturbations\cite{bhbook, shen_book, hasan-kane_TI, kane-mele, fu-kane-mele}. The topological invariants associated with the boundary modes have been studied in diverse contexts\cite{tknn, bellissard-baldes_noncomm_IQHE, baldes_disorder_IQHE, prodan_disorder_TI, volovik_book, essin-gurarie_bulk_bdry}, and related to the topological invariants of the bulk under the \emph{bulk-boundary correspondence}.

Many of these ideas have also been generalized to semimetals, which are gapped almost everywhere in the bulk Brillouin zone(BZ). The intuition for insulators can thus often be carried over, albeit with momenta defined only on a punctured BZ. A notable example are the 3+1 dimensional \emph{Weyl semimetals}(WSMs)\cite{turner-ashvin2011, witten_wsm}, gapless at a finite (even) number of isolated, topologically protected \emph{Weyl nodes} in the bulk BZ. The surface exhibits \emph{Fermi arcs}, curves in the surface BZ of topologically protected zero energy modes connecting the projections of the Weyl nodes. Recently, WSMs have attracted much attention recently owing to various experimental realizations\cite{hasan_wsm, hasan_wsm2, hasan_wsm3, marin_wsm_photonic,marin_wsm_photonic2}.  

WSMs are particularly interesting because the \emph{nodal configuration}, i.e, the positions and chiralities of Weyl nodes (or equivalently, the linearized theory of the bulk) does not completely specify the surface spectrum for a fixed boundary condition. Instead, one also needs to know the \emph{Fermi arc connectivity}, i.e, the pairs of nodes that are connected by Fermi arcs on a given surface, which may depend on the microscopic details of the bulk\cite{vd-str_wsm}. Mathematically, the connectivity can be described in terms of a relative homology group\cite{mathai_wsm_top} or \emph{Euler structures}\cite{mathai_semimetal_top}, which encode the additional global information required to completely characterize a WSM.

The arguments for existence of boundary modes in topological phases can be readily generalized to \emph{interface modes} localized at junctions between topological phases, where one would expect localized modes whenever the two phases forming the junction are topologically distinct\footnote{
  Indeed, the boundary modes for topological phases can be thought of as a special case corresponding to one of these phases being the vacuum, a topologically trivial insulator with infinite energy gap.
}. 
Since the formation of a junction can be intuitively visualized as the coupling (due to proximity) of the surface modes of the two topological phases involved, the interface modes can be expected to inherit aspects of topology from both phases. Various such cases involving topological insulators (TIs) and superconductors (TSCs) have already been studied, notable examples being \mbox{TI-TI} junctions\cite{takahashi_ti_opp_vel, beule_ti_junc}, \mbox{TI-WSM} junctions\cite{adolfo_ti_wsm,juergens_ti_wsm} and \mbox{WSM-TSC} junctions\cite{sumathi_wsm_sc, chen_wsm_sc}. 

\begin{figure}
	\centering
	\includegraphics[width=0.85\columnwidth]{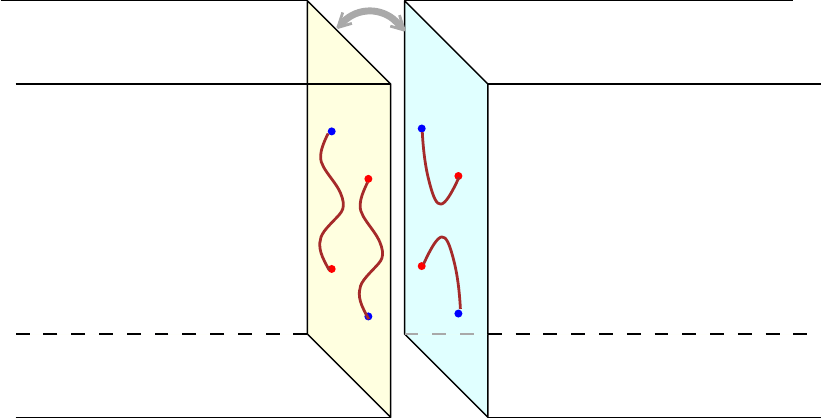}
	\caption{
	  A schematic depiction of an interface between WSMs with different Fermi arc connectivities. The hopping between the two surfaces, denoted by a gray arrow, leads to hybridization of the localized modes on the two surfaces, i.e, a \emph{Fermi arc reconstruction}.
	}
	\label{fig:sch}
\end{figure}

In this article, we consider the interface modes for a WSM-WSM junction. For two WSMs with different nodal configurations, one generically expects zero modes corresponding to the ``difference'' between the Fermi arcs on the two sides. However, a more intriguing possibility is to consider two WSMs with identical nodal configurations, but different Fermi arc connectivities, as schematically illustrated in \fig{fig:sch}. These Fermi arcs cannot be coupled and hence gapped out by hopping terms between surface modes that maintain the translation invariance in the transverse directions, since they lie on different transverse momenta(in the surface BZ). 

We find that the coupling between the Fermi arcs of the two surfaces leads to a \emph{Fermi arc reconstruction}. In the tight binding models considered here, the Fermi arcs generically combine to form closed loops of zero modes localized at the interface. In certain cases, these loops detach from the Weyl nodes, leading to a stable Fermi surface of effectively 2D modes. Given tight-binding models for WSMs, these modes can be obtained by a numerical exact diagonalization(ED) of the real space Hamiltonian. 

To study the Fermi arc reconstruction analytically, we generalize the transfer matrix method to include interfaces. This approach, being partly a real space method, is ideally suited to probing spatially inhomogeneous systems. 
It has also been used to analytically study the bulk and boundary states of noninteracting toplogical phases\cite{hatsugai_cbs, hatsugai_cbs_PRL, dh_lee_cbs, schulman_cbs, tauber-delplace_tm, vd-vc_tm}. Building up on this literature, we derive a closed form condition for the existence of interface modes in terms of the transfer matrices for systems on the two sides of the interface, which is applicable to interfaces between systems described by tight-binding models. 

The rest of this article is organized as follows: In Sec.~\ref{sec:wsm}, a generic two band tight-binding model for WSMs is introduced and the associated transfer matrix is computed. In Sec.~\ref{sec:int}, general conditions for the presence of interface modes are derived, which are then applied to specific WSM models in Sec.~\ref{sec:eg}. We conclude with a discussion in Sec.~\ref{sec:conc}, and some of the tedious calculations are relegated to the appendices.

\section{Weyl semimetals}     \label{sec:wsm}
In this section, we study a general family of lattice models which can realize a WSM with an arbitrary (even) number of Weyl nodes lying in a plane in the bulk BZ. We construct the transfer matrix for translations normal to this plane and use it to study the surface modes.

\subsection{The lattice model}   \label{sec:wsm_model}
Consider the family of 2-band models described by the Bloch Hamiltonian
\beq 
  \hlt(\vk) = \hlt_x(k_x) + \hlt_\perp(\vk_\perp),   \label{eq:hlt_orig} 
\eeq 
where we take
\begin{align} \label{eq:4nodewsm1}
  \hlt_x(k_x) = & \; \sin k_x \sigma^x + (1 - \cos k_x) \sigma^z, \nonumber \\ 
  \hlt_\perp(\vk_\perp) = & \; \hlts_y(\vk_\perp) \sigma^y + \hlts_z(\vk_\perp) \sigma^z. 
\end{align}
Here, $\bet = \left( \hlts_y, \hlts_z \right)^T : \torus^2 \to \real^2$ are functions of the transverse momentum $\vk_\perp \equiv (k_y, k_z) \in \torus^2$, and the Pauli matrices describe the internal (\emph{pseudospin}) degrees of freedom. The spectrum is gapless in the $k_x = 0$ plane for 
\beq 
  \hlts_y(\vk_\perp^\ast) = \hlts_z(\vk_\perp^\ast) =  0.   \label{eq:nodes}
\eeq 
We preclude the possibility of gapless modes in the $k_x = \pi$ plane by demanding that $\hlts_z > -2$. 

The Hamiltonian can be linearized near a particular gapless point $\vk^\ast = (0, \vk_\perp^\ast)$ as 
\[
 \hlt \approx \delta k_a V_{ab} \sigma^b, \quad a,b \in \{x, y,z\},
\]
where 
\beq 
  V = \begin{pmatrix}
	1 & 0 \\ 
	0 & \jac 
      \end{pmatrix};  
      \quad \jac(\vk_\perp^\ast) = \left. \frac{\partial\bet}{\partial\vk_\perp} \right|_{\vk_\perp^\ast}
\eeq 
being the Jacobian of $\bet$. The system exhibits a Weyl node at $\vk^\ast$ iff $V$ (and hence $\jac$) is nonsingular, in which case the associated chirality is given by 
\beq 
  \chi = \sgn{ \det{V(\vk^\ast)} } = \sgn{ \det{\jac(\vk_\perp^\ast)} }.
\eeq 

Choosing a suitable $\bet$, WSMs with arbitrary (even) number of Weyl nodes can be realized. We expect Fermi arcs on a surface normal to $x$, with the surface BZ parametrized by $\vk_\perp$ and the surface projections of the Weyl nodes simply the solutions of \eq{eq:nodes}. To study the surface spectrum analytically, we next construct the transfer matrix for translations along the $x$-direction.

\subsection{Constructing the transfer matrix}    \label{sec:wsm_tm}
Consider the model of \eq{eq:hlt_orig} on a cylinder $[0,N] \times \torus^2$, finite along $x$. Since $k_x$ is not a good quantum number anymore, inverse Fourier transform the Hamiltonian of \eq{eq:hlt_orig} along $x$ to get 
\begin{align}
  \hlt_{\text{real}}(\vk_\perp) = & \; \sum_{n}  \bigg[  \vcd_{n+1} \left( \frac{ \sigma^x - i \sigma^z}{2} \right) \vc_n + \hc \nonumber \\ 
  & + \vcd_n \left( \hlts_y \sigma^y + (1 + \hlts_z) \sigma^z \right) \vc_n \bigg],    \label{eq:hlt_real}
\end{align}
where $\vc_n (\vk_\perp) = (\c_{n,1}, \c_{n,2})^T$ are the annihilation operators for the two pseudospin states at the layer $n$ along $x$. 
The Schr\"odinger equation, $\hlt \ket{\Psi} = \ve \ket{\Psi}$, can be written as a recursion relation\cite{hatsugai_cbs, vd-vc_tm}
\beq 
  J \Psi_{n+1} + M \Psi_n + J^\dg \Psi_{n-1} = \ve \Psi_n,    \label{eq:recur}
\eeq 
where $\Psi_{n} \in \cmplx^2$ 
and 
\begin{align}
  J = \frac{1}{2i} \left( \sigma^x - i \sigma^z \right), \quad   M = \hlts_y \sigma^y + (1 + \hlts_z) \sigma^z
\end{align}
are the \emph{hopping} and \emph{on-site} matrices, respectively. 

The conventional transfer matrix construction works by rewriting \eq{eq:recur} as an equation for $\Psi_{n+1}$ in terms of $\Psi_{n}$ and $\Psi_{n-1}$ by inverting $J$. However, since $J$ is singular ($\rank J = 1$), we follow the alternative construction introduced in Ref~\onlinecite{vd-vc_tm}, where the recursion relation is instead rewritten as
\beq 
  \Psi_n = \green \left( J \Psi_{n+1} + J^\dg \Psi_{n-1} \right),   \label{eq:recur_G}
\eeq 
where 
\begin{align}
  \green \equiv & \;  (\ve \id - M)^{-1} \nonumber \\ 
  = & \; \frac{1}{\ve^2 - \hltsq^2} \left[ \ve \id + \hlts_y \sigma^y + (1 + \hlts_z) \sigma^z  \right],
\end{align}
can be identified as an \emph{on-site Green's function}, with 
\beq 
  \hltsq^2 = \hlts_y^2 + (1 + \hlts_z)^2.  \label{eq:gamma_def}
\eeq
Next, we take the reduced singular value decomposition of $J$, \viz, $J = \vv \cdot \vw^\dg$, where 
\beq 
  \vv = \frac{1}{\sqrt{2}} \vecenv{-i}{1}, \quad  
  \vw =  \frac{1}{\sqrt{2}} \vecenv{i}{1}, 
\eeq 
which satisfy
\[ 
  \vv^\dg \vv = \vw^\dg \vw = 1, \quad \vv^\dg \vw = 0,
\] 
since $J^2 = 0$. Thus, $\{\vv, \vw\}$ forms an orthonormal basis of $\cmplx^2 \ni \Psi_n$, and we may expand $\Psi_n$ as  
\begin{align} 
  \Psi_n = & \; \left( \vv^\dg \Psi_n \right) \vv + \left( \vw^\dg \Psi_n \right) \vw  \nonumber \\ 
  \equiv & \;  \alpha_n \vv + \beta_n \vw, 
\end{align}
The recursion relation of \eq{eq:recur_G} thus becomes 
\beq 
  \Psi_n = \green \vv \beta_{n+1} + \green \vw \alpha_{n-1},  
\eeq 
so that the coefficients in the $\{\vv, \vw\}$ basis are 
\begin{align}
 \alpha_n = & \; \vv^\dg \green \vv \, \beta_{n+1} + \vv^\dg \green \vw \, \alpha_{n-1},  \nonumber \\  
 \beta_n = & \; \vw^\dg \green \vv \, \beta_{n+1} + \vw^\dg \green \vw \, \alpha_{n-1},
\end{align}
where 
\begin{align}
 & \vv^\dg \green \vv = \frac{\ve  + \hlts_y}{\ve^2 - \hltsq^2}, \quad \vw^\dg \green \vw = \frac{\ve - \hlts_y}{\ve^2 - \hltsq^2}, \nonumber \\
 & \vv^\dg \green \vw = \vw^\dg \green \vv = - \frac{1 + \hlts_z}{\ve^2 - \hltsq^2}.
\end{align}
These equations can finally be rearranged to get a transfer matrix equation:
\beq 
  \Phi_{n+1} = T \Phi_n, \quad \Phi_n \equiv \vecenv{\beta_n}{\alpha_{n-1}},
\eeq 
where the transfer matrix is
\beq 
  T(\ve, \vk_\perp) = \frac{1}{1 + \hlts_z}\left( 
  \begin{array}{cc} 
    \ve^2 - \hltsq^2 & \quad  -(\ve - \hlts_y) \\ 
    \ve + \hlts_y & \quad -1    
  \end{array} \right).    \label{eq:tmat}
\eeq 
Note that the vectors $\Phi_n$ that are actually translated by the transfer matrix consist of the physical wavefunctions at two adjacent sites.

\begin{figure}[b!]
	\centering
	\includegraphics[width=0.65\columnwidth]{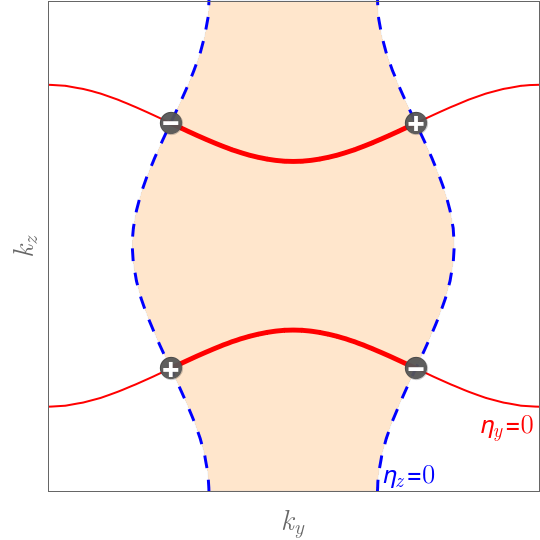}
	\caption{ 
	  The geometric picture of the Weyl nodes and Fermi arcs for the tight-binding model of \eq{eq:hlt_orig} (with $\bet$ defined by \eq{eq:4nodewsm} for $\varphi = 4\pi/3$). The Weyl nodes occur in the the $k_x = 0$ plane at the intersection of the curves defined by $\hlts_y = 0$ and $\hlts_z = 0$, depicted by solid red lines and dashed blue lines, respectively. The surface has localized modes for $\vk_\perp = (k_y, k_z)$ in the shaded region, and the subset of these modes at zero energy, viz, the Fermi arcs, are denoted by the bold red lines. 
	}
	\label{fig:wsm_engg_sch}
\end{figure}

\subsection{Surface states}    \label{sec:wsm_surf}
The spectrum of the transfer matrix (for a given $\ve$ and $\vk_\perp$) encodes the bulk/edge states of the system. More explicitly, note that the transfer matrix can be used to write wavefunctions as $\Phi_n = T^n \Phi_0$, so that knowing $\Phi_0$ completely describes a wavefunction. It follows that given an eigenvector,
\beq 
  T \Phi_0 = \rho \Phi_0 \implies \Phi_n = \rho^n \Phi_0,
\eeq 
so that $\Phi$ describes a bulk (Bloch) state if $\abs{\rho}=1$ and an edge state exponentially localized at the left(right) edge for $\abs{\rho}<1$ ($\abs{\rho}>1$). Furthermore, the boundary conditions are used to obtain the allowed $\Phi_0$'s for the physical eigenstates of the Hamiltonian. 

For surface states, one typically considers the Dirichlet boundary condition ($\Psi_{-1} = \Psi_{N+1} = 0$), which translates to demanding that
\beq 
  \Phi_0 \propto \vecenv{1}{0}, \quad 
  \Phi_{N+1} \propto \vecenv{0}{1}, \quad 
\eeq 	
at the left and the right edge, respectively. For the left edge, we explicitly compute 
\beq 
  T \Phi_0 = \rho \Phi_0 \implies 
  \begin{cases} 
    \ve = -\hlts_y & \\ 
    \rho = -(1 + \hlts_z). & 
  \end{cases}
\eeq 
Thus, for a given $\vk_\perp$, there exists a surface mode with energy $\ve = -\hlts_y(\vk_\perp)$ exponentially localized\footnote{ 
  We have assumed a semi-infinite system. 
} at the left surface if $\abs{\rho}<1$, i.e, if
\beq 
  -1 \leq 1 + \hlts_z(\vk_\perp) \leq 1 \implies \hlts_z(\vk_\perp) \leq 0, 
\eeq 
since we have assumed that $\hlts_z > -2$. Finally, for the left surface, the Fermi arc, i.e, the locus of the surface modes with $\ve = 0$, are given by 
\beq 
  \hlts_y(\vk_\perp) = 0, \quad   \hlts_z(\vk_\perp) < 0,    \label{eq:fermi_arc}
\eeq
with the surface projection of the Weyl nodes given by the end points $\hlts_y(\vk_\perp) = 0 = \hlts_z(\vk_\perp)$. An identical calculation for the right surface (i.e, using $\Phi_{N+1}$) yields surface modes dispersing in the opposite direction, i.e, $\ve = \hlts_y$. The corresponding Fermi arcs are given by 
\beq 
  \hlts_y(\vk_\perp) = 0, \quad   \hlts_z(\vk_\perp) < 0,    \label{eq:fermi_arcR}
\eeq
which is identical to those on the left surface. 

Eqns \ref{eq:nodes} and \ref{eq:fermi_arc} have a straightforward geometric interpretation, which provides useful intuition for the construction of WSM lattice models. The level sets $\hlts_y = 0$ and $\hlts_z = 0$ generically define two sets of curves on $\torus^2$, whose intersections are the Weyl nodes (in the $k_x = 0$ plane). The Fermi arcs then lie along the former ($\hlts_y = 0$) curves, stretching between the intersections with the latter curve. A particular example is shown in \fig{fig:wsm_engg_sch}.


\section{Interfaces}    \label{sec:int}
In this section, we consider interfaces between noninteracting topological phases described by tight binding models, and derive a general analytical condition for the existence of a localized mode at the interface in terms of the associated transfer matrices. We apply these conditions to the transfer matrix derived in \sect{sec:wsm} for simple interfaces where the analytical solutions are particularly illuminating.

\subsection{A qualitative picture}  \label{sec:int_qual}
Before we get into explicit computations, we seek to qualitatively understand the circumstances under which one might get modes localized at an interface between two topological phases. To arrive at an intuitive picture, consider the two phases being completely decoupled, with their individual surface modes with energies in the corresponding bulk gaps. As one turns on a hopping between the two surfaces (hereafter referred to as \emph{gluing}), these modes may gap out in pairs. Since the interface does not break the translation symmetry in the transverse directions, the corresponding momenta are conserved, so that only hoppings between modes with identical $\vk_\perp$ are allowed. 

Explicitly, let there be localized modes with energies $\ve_{1,2} (\vk_\perp)$ on the two surfaces before they are glued. Then, the most general effective Hamiltonian for the interface can be written as 
\beq 
  \hlt_{\text{eff}}(\vk_\perp) = 
  \begin{pmatrix}
    \ve_1(\vk_\perp) & t(\vk_\perp) \\ 
    t^\ast(\vk_\perp) & \ve_2(\vk_\perp),
  \end{pmatrix}
\eeq 
whose spectrum is
\[
  \ve = \frac{1}{2} \left( \ve_1 + \ve_2 \pm \sqrt{(\ve_1 - \ve_2)^2 + 4 \abs{t}^2} \right).
\]
The locus of zero energy modes is then given by 
\beq
  (\ve_1 + \ve_2)^2 -(\ve_1 - \ve_2)^2 =  4 \abs{t}^2.       \label{eq:glued_zero_modes}
\eeq
Given $t(\vk_\perp)$, this equation describes the zero energy modes localized at the interface. 

One can immediately consider one trivial case, \viz, when $\ve_1 = -\ve_2$, for which there are no zero modes for any choice of $t$. This is the case, for instance, if the two systems being glued are identical, since the left and right surface states being glued have surface states that disperse in the opposite directions. This is to be expected, since after gluing one is left with a single uniform system, which does not have any localized modes in the bulk. 

While good to obtain some intuition, this picture is usually not complete, since the relation between $t(\vk_\perp)$ and the actual boundary conditions at the interface are not obvious. Furthermore, there is no direct way to study the associated wavefunction, to compute, for instance, the penetration depth, without resorting to a real space ED calculation. Therefore, in the next subsection, we shall instead use the transfer matrix approach to study the interface modes.

In the case of WSMs, one is usually interested in the zero energy surface modes, \viz, Fermi arcs. At low energies, the corresponding linearized spectrum is that of a chiral fermion dispersing in the direction normal to the arc\footnote{
  Recall that since a WSM can be thought of as stacked Chern insulators, each point of the Fermi arc is the zero crossing of the edge spectrum of a Chern insulator. 
}. 
Let the two surfaces have Fermi arcs with normal vectors(lying in the surface BZ) $\vn_{1,2}$ near some $\vk_\perp$, so that in a neighborhood, the associated linearized spectra are $\ve_i = \delta \vk_\perp \cdot \vn_i$. Using \eq{eq:glued_zero_modes}, the zero modes after gluing are given by 
\beq 
  \left[ \delta\vk_\perp \cdot (\vn_1 + \vn_2) \right]^2 - \left[ \delta\vk_\perp \cdot (\vn_1 - \vn_2) \right]^2 = 4 \abs{t}^2.
\eeq
Generically, this is an equation for a hyperbola, which can be interpreted as a \emph{level repulsion}, albeit in the $\vk_\perp$ space. One extreme case is to have $\vn_1 = \vn_2$, corresponding to parallel Fermi arcs running in the same directions. The interface modes are then given by  $\left( \delta\vk_\perp \cdot \vn_1 \right)^2 = \abs{t}^2$, which denotes two lines separated by $2 \abs{t}$. We shall revisit this case in \sect{sec:eg_2}. 

Finally, we conclude with a conundrum for WSMs: Consider an interface between two WSMs with identical nodal configuration, for which surface modes exist in disjoint regions of the surface BZ. In general, one would not expect modes at the interface, since the two WSMs are topologically identical. However, the gluing mechanism described above precludes gapping out the original Fermi arcs, since there are no localized modes at the same $\vk_\perp$ to couple to.  We study this case in more detail in \sect{sec:eg_4}, and discover that there are, in fact, modes localized at the interface.

\begin{figure*}
	\centering
	\includegraphics[ width=0.95\textwidth]{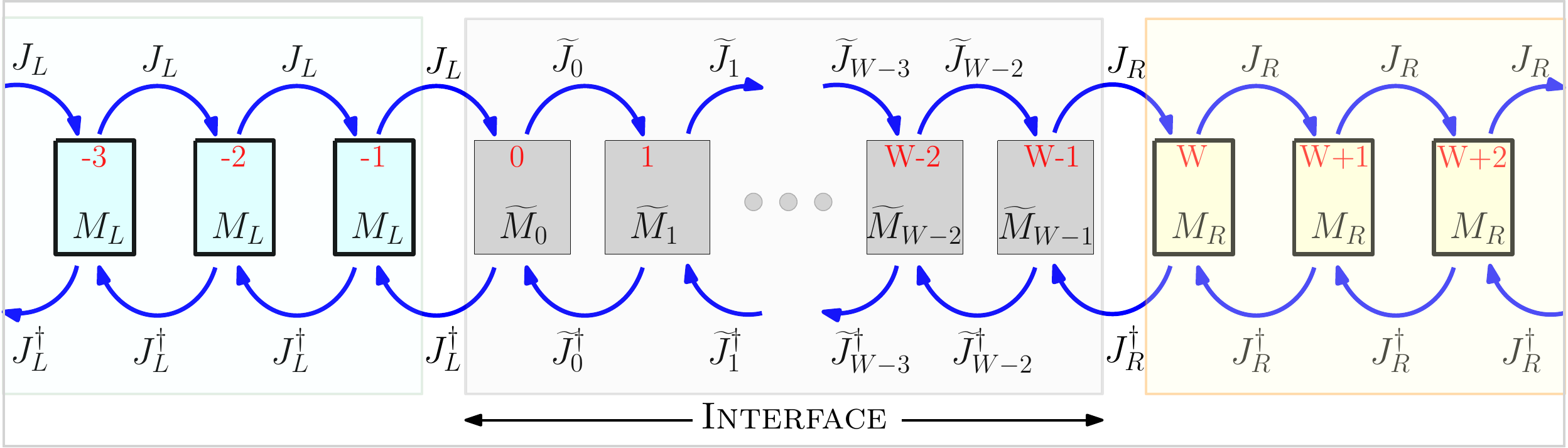}
	\caption{ 
	  The basic setup for describing the interface in the transfer matrix formalism. The left and right bulk layers are depicted by green and yellow, while the interface layers are depicted in gray. The matrices $J$'s and $M$'s are generically functions of the transverse momentum. 
	}
	\label{fig:int}
\end{figure*}

\subsection{The general condition}     \label{sec:int_gen}
The transfer matrix approach, being inherently a real space method, is ideally suited for the analytical study of interfaces. In particular, the transfer matrix description describes the boundary modes as the solution of two conditions, \viz, the \emph{decay condition}, associated with the eigenvalues of the transfer matrix and the \emph{boundary condition} corresponding to the associated eigenvectors. The physical states are naturally required to satisfy both, but choosing either one separately, one may often extract additional topological information about the system\cite{vd-vc_tm}. 

We now adapt this approach to study the interface modes. As the interface naturally breaks the translation invariance in the $x$ direction, consider the transfer matrices translating along $x$ for the phases to the left ($T_L$) and right ($T_R$) of the interface. Furthermore, there can generally be additional on-site terms at the interface, so we assume that the interface is $W\geq 0$ layers wide. Setting $n=0$ to be its left end, we can use the transfer matrices to write the wavefunctions for $n \notin [1,W-1]$ as 
\beq 
  \Phi_n = \begin{cases}
             T_R^{n-W} \Phi_W, & n \geq W,  \\ 
             T_L^{n} \Phi_0, & n \leq 0.    \label{eq:tmatLR}
           \end{cases}
\eeq 
Thus, a mode localized at the interface must satisfy
\begin{align}
 & T_R \Phi_W = \rho_R \Phi_W,  & \!\!\!\!\!\!\!\!\!\!\!\!\!\!\!\! \abs{\rho_R} < 1,  &  \quad \nonumber \\ 
 & T_L \Phi_0 = \rho_L \Phi_0,  & \!\!\!\!\!\!\!\!\!\!\!\!\!\!\!\! \abs{\rho_L} > 1.  &  \quad   \label{eq:rhoLR_cond}
\end{align}
Furthermore, let the effect of the interface on a wavefunction be encoded in an invertible matrix $S$, so that 
\beq 
 \Phi_W = S \Phi_0.   \label{eq:S_def}
\eeq 
Physically, this matrix can be thought of as the transmission matrix for a scattering problem with the interface being the scatterer. This is inherently a real space quantity, and depends on the geometry of the interface and the on-site potential(s). In the next subsection, we shall compute it explicitly for various cases. 

The conditions for the existence of a localized interface mode can now be separated into a \emph{matching} and a \emph{decay} condition. This is a discrete analogue of the strategy for computing the bound states of a 1D potential well in quantum mechanics. To wit, one obtains general solutions (with certain unknown constants) to the left and the right of the potential well which exhibit the desired asymptotic behavior at $\pm \infty$, and then fixes the constants by \emph{matching} the solutions across the potential well. We shall follow a similar strategy here, however, we shall find unlike the case of potential wells, the matching conditions are easier to compute in terms of the transfer matrices. 


We note that using Eq.~\ref{eq:tmatLR} and \ref{eq:S_def}, a wavefunction of the system is completely determined by the knowledge of $\Phi_0$. Thus, we shall seek to derive a condition only in terms of $\Phi_0$. In particular, \eq{eq:rhoLR_cond} reduces to
\beq 
  S^{-1} T_R S \Phi_0 = \rho_R \Phi_0, \quad T_L \Phi_0 = \rho_L \Phi_0,
\eeq 
alongwith the condition on $\rho_{L,R}$. 

For the matching condition, we ignore the constraints on $\rho_{L,R}$ and simply demand that $\Phi_0 \neq 0$ be a simultaneous eigenvector of $S^{-1} T_R S$ and $T_L$, which implies that 
\beq 
  [T_L, S^{-1} T_R S]\Phi_0 = 0.    \label{eq:cond_comm}
\eeq  
As it turns out\cite{shemesh_common_eigs}, for a $2\times 2$ transfer matrix\footnote{ 
  For larger transfer matrices, the latter condition is necessary, but not sufficient for a shared eigenvector. A more general condition for arbitrarily-sized transfer matrices is derived in Appendix \ref{app:cond}. 
}, this is a necessary as well as \emph{sufficient} condition for the existence of a simultaneous eigenvector. Finally, \eq{eq:cond_comm} has a nontrivial solution for $\Phi_0$ iff the commutator satisfies the Cramers' condition:
\beq 
  \det\left( [T_L, S^{-1} T_R S] \right) = 0.   \label{eq:cond_det}
\eeq 
The LHS is a function of $\ve$ and $\vk_\perp$, so that we get an implicit equation for the points that satisfy the matching condition. 



Given a particular $(\ve, \vk_\perp)$ satisfying \eq{eq:cond_det}, one can then compute $\Phi_0 \in \ker \left( [T_L, S^{-1} T_R S] \right)$, so that $\Phi_0$ is an eigenvector of $T_L$ while $S \Phi_0$ is an eigenvector of $T_R$. For the corresponding eigenvalues, we have the following three possibilities: 
\begin{enumerate}
 \item $\abs{\rho_R} < 1, \, \abs{\rho_L} > 1$.  
 \item $\abs{\rho_R} > 1, \, \abs{\rho_L} < 1$.  
 \item $\abs{\rho_A} \lessgtr 1$ for both $A = L,R$.  
 \label{blah}
\end{enumerate}
The first case corresponds to a mode decaying into the bulk on both sides of the interface, i.e, the (\emph{physical}) decay condition. The remaining cases are also relevant, since they can be thought of as the physical interface modes for a different choice of the gluing matrix $S$. For instance, the second case, with a mode growing in the bulk on both sides, can be obtained by swapping the two phases ($L \leftrightarrow R$) and replacing the gluing matrix $S$ by $S^{-1}$. The relevance of these ``unphysical'' cases is discussed in more detail in Appendix \ref{app:unphys}. 

Finally, we note that the eigenvalues $\rho_{L,R}$ must vary continuously with $\vk_\perp$ and cross the unit circle only in pairs\cite{vd-vc_tm}, these labels cannot change unless the curves intersect\footnote{
  Parts of these ``curves'' can actually be single points. This is the case, for instance, for the projections of the Weyl nodes on the interface BZ.
}. We can then label each segment of the curves obtained from the matching conditions by the eigenvalues $\rho_{L,R}$. Thus, in practice, given a pair of phases described by $T_L$ and $T_R$ and a gluing matrix $S$, we shall compute the curves given by the matching condition (\eq{eq:cond_det}) in the interface BZ. For each section of the curve,  we then compute the null eigenvector of $[T_L, S^{-1}T_R S]$ to assign it one of the three possibilities discussed above. 



\subsection{Computations for specific interfaces}     \label{sec:int_spc}
The conditions for interface modes derived in the previous subsection do not refer to the specific form of the transfer matrices or interfaces. We now consider the specific case of the transfer matrix derived in \sect{sec:wsm}. We begin by recalling that for a translation invariant system, the Schr\"odinger equation can be written as a recursion relation\cite{vd-vc_tm}, as shown in \eq{eq:recur}. For spatially inhomogeneous case, this generalizes to
\beq 
  J_n \Psi_{n+1} + M_n \Psi_n + J_{n-1}^\dagger \Psi_{n-1} = \ve \Psi_n,       \label{eq:schr_pos_dep}
\eeq
where both the hopping and on-site matrices are position dependent. Consider the setup depicted in \fig{fig:int}, which explicitly corresponds to setting 
\begin{align}
  J_n = & \; \begin{cases} 
	  J_L, & n \leq -1, \\ 
	  \wt{J}_n, & 0 \leq n \leq W-2, \\ 
	  J_R, & n \geq W-1,   
        \end{cases} 
  \nonumber \\ 
  M_n = & \; \begin{cases} 
	  M_L, & n \leq -1, \\ 
	  \wt{M}_n, & 0 \leq n \leq W-1, \\ 
	  M_R, & n \geq W,    
        \end{cases}
  \label{eq:}
\end{align} 
where $\wt{J}_{n}$ and $\wt{M}_{n}$ encode the interface of width $W \geq 0$. The case with $W=0$ is not well defined unless $J_R = J_L$. So far, this setup is valid for interfaces between any systems described by tight-binding models.

In the rest of this article, we restrict ourselves to interfaces between two WSMs described by \eq{eq:hlt_orig} with $\bet = \bet^{L/R}$ to the left/right of the interface, so that $J_L = J_R = J$ and 
\beq 
  M_A = \hlts_y^A \sigma^y + (1 + \hlts_z^A) \sigma^z, \quad A = L, R.
\eeq 
The transfer matrices $T_{L,R}$ for the two sides were computed in \sect{sec:wsm_tm}. We now seek to apply the results of \sect{sec:int_gen} to study the zero modes, if any, localized at the interfaces. The simplest case of an interface is is an abrupt one, with no additional local potential. We shall keep the coupling between the two sides to be tunable, so that the hopping matrix connecting the two sides is simply $\kappa J$, with $\kappa \in [0,\infty)$, the two limits corresponding to the uncoupled and the strongly coupled case. A quick glance at \fig{fig:int} reveals that for $\wt{J}_0 = \kappa J$, we must take $W = 2$; however, for $\kappa = 1$, we may take $W = 0$, rendering it particularly simple to analyze. This is the case that we start with.



\begin{figure}[t]
	\centering
	\includegraphics[ width=0.7\columnwidth]{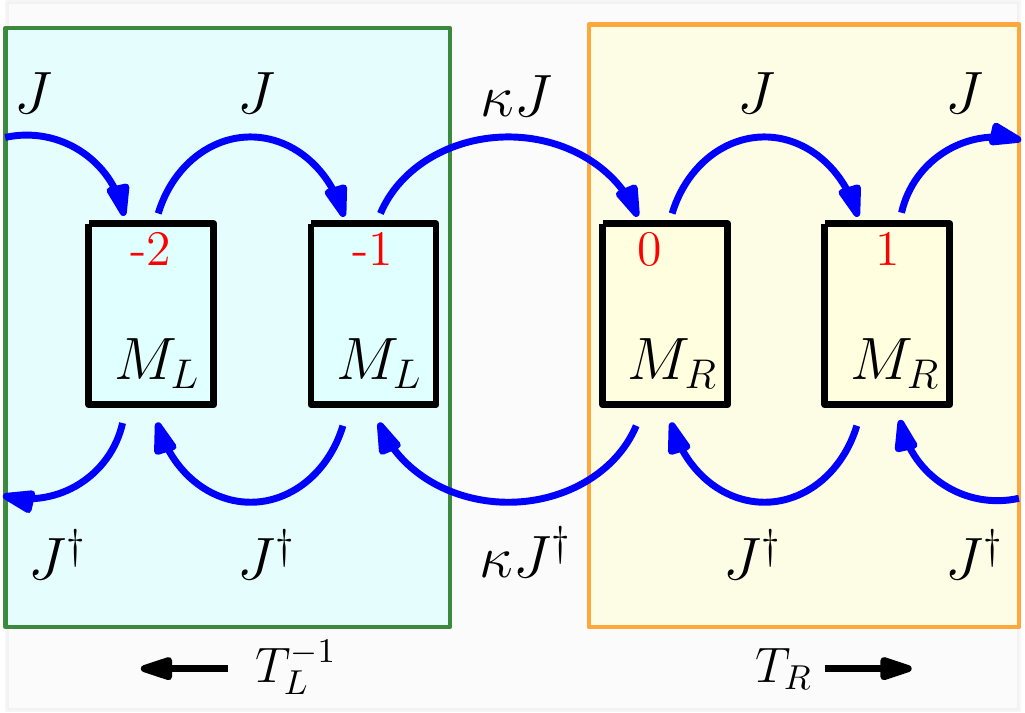} \\ \ \\ 
	\includegraphics[ width=0.8\columnwidth]{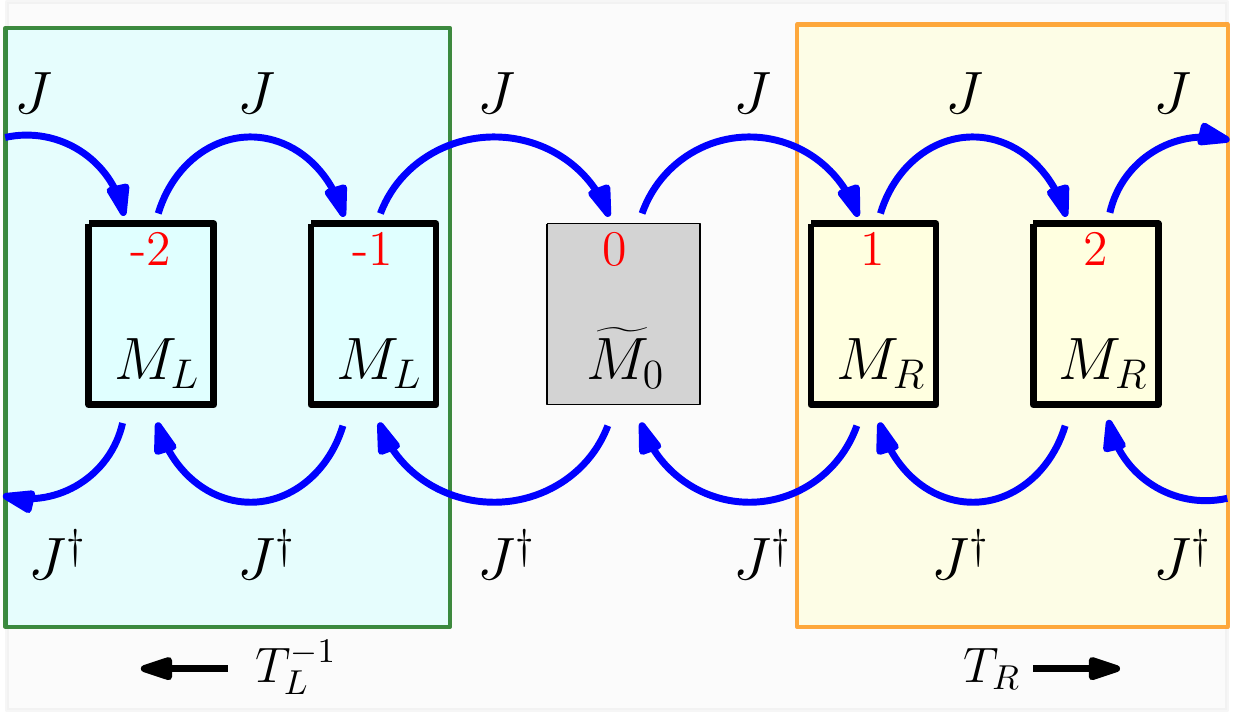}
	\caption{ 
	  Schematic of the interfaces discussed in \sect{sec:int_spc}: (Top) A hard interface with a tunable coupling $\kappa$ between the surfaces. (Bottom) A softer interface with an on-site potential described by $\wt{M}_0$.
	}
	\label{fig:int_eg}
\end{figure}

\subsubsection{Hard interface for $\kappa = 1$} 
We set $W = 0$, so that 
\[
  J_n = J, \quad 
  M_n = \begin{cases} 
	  M_L, & n \leq -1, \\
	  M_R, & n \geq 0.    
        \end{cases}  
\]
The matrix $S$ is required to satisfy $S\Phi_0 = \Phi_W$, so that we need $S = \id$. The matching condition for the interface then becomes 
\beq 
  \det\left( [T_L, T_R] \right) = 0.
\eeq 
Using the transfer matrices $T_{L,R}$ computed in \eq{eq:tmat}, we show in Appendix \ref{app:calc} that 
\begin{align} 
   \det\left( [T_L, T_R] \right) = & \; \det\left(  a \, \ve \, \sigma^x + i (b \, \ve^2 + c) \, \sigma^y + d \,\ve \, \sigma^z \right)  \nonumber \\ 
  = & \; a^2 \ve^2 - (b \ve^2 + c)^2 + d^2 \ve^2,     \label{eq:int_cond}
\end{align}
where 
\begin{align}
 a = & \; \hltsq_L^2 - \hltsq_R^2, \nonumber \\  
 b = & \; \hlts_y^R - \hlts_y^L,   \nonumber \\ 
 c = & \; \hlts_y^R \left\{ 1 - \hltsq_L^2  \right\} - \hlts_y^L \left\{ 1 - \hltsq_R^2  \right\},  \nonumber \\ 
 d = & \; 2 \left( \hlts_y^R - \hlts_y^L \right),
\end{align}
with $\hltsq$ as defined in \eq{eq:gamma_def}. 

For zero modes we set $\ve = 0$, and the matching condition reduces to $c=0$. 
A trivial solution to that condition is $\hlts_y^R = \hlts_y^R = 0$, which corresponds to overlapping Fermi arcs for the two side. For $\hlts_y^R, \hlts_y^R \neq 0$, the matching condition can be expressed elegantly as
\beq 
 \frac{1 - \hltsq_L^2}{\hlts_y^L} = \frac{ 1 - \hltsq_R^2 }{\hlts_y^R}.   \label{eq:cond_w0}
\eeq 
Each of these is a solution of $\det[T_L, T_R] = c^2 = 0$ with multiplicity two. Tuning away from $\kappa=1$ splits these solutions apart, as we discuss next. 


\subsubsection{Hard interface for arbitrary $\kappa$}   \label{sec:int_eg_2}
\newcommand{\kmat}{K}
We set $W=2$ and 
\[
  \wt{M}_0 = M_L, \quad \wt{M}_1 = M_R, \quad \wt{J}_0 = \kappa J,
\]
so that explicitly,
\[
  J_n = \begin{cases} 
	  \kappa J, & n = 0, \\
	  J, & n \neq 0,    
        \end{cases}, \quad   
  M_n = \begin{cases} 
	  M_L, & n \leq 0, \\
	  M_R, & n \geq 1,    
        \end{cases}.
\] 
The matrix $S$ is required to satisfy $S\Phi_0 = \Phi_2$. To compute it, consider the recursion relations for $n=0,1$ (involving $\kappa$), which can be explicitly written as 
\begin{align*}
  \kappa J \Psi_1 + M_L \Psi_0 + J^\dagger \Psi_{-1} = \ve \Psi_0,    \nonumber \\  
  J \Psi_2 + M_R \Psi_1 + \kappa J^\dagger \Psi_0 = \ve \Psi_1.
\end{align*}
These can be solved by a construction similar to that discussed in \ref{sec:wsm}. Explicitly, for $n=0$, we get 
\begin{align}
 \alpha_0 = & \; \kappa \vv^\dg \green_L \vv \, \beta_{1} + \vv^\dg \green_L \vw \, \alpha_{-1},  \nonumber \\  
 \beta_0 = & \; \kappa \vw^\dg \green_L \vv \, \beta_{1} + \vw^\dg \green_L \vw \, \alpha_{-1}, 
\end{align}
where $\green_A = \left( \ve \id - M_A \right)^{-1}, \; A = L,R$. We get a similar expression for $n=1$, and constructing the transfer matrices, we get
\begin{align}
 \Phi_1 & \; = \diag{\frac{1}{\kappa}, 1} \cdot T_L \Phi_0, \nonumber \\ 
 \Phi_2 & \; = T_R \cdot \diag{1, \kappa} \Phi_1,
\end{align}
so that
\beq 
  S = T_R \kmat T_L, \quad \kmat = \diag{\frac{1}{\kappa}, \kappa}.
\eeq
The condition derived in \eq{eq:cond_det} thus simplifies to
\beq 
  \det\left( [T_L, \kmat^{-1} T_R \kmat] \right) = 0.    \label{eq:cond_w2}
\eeq
For the transfer matrices of \eq{eq:tmat}, we simplify \eq{eq:cond_w2} using Mathematica\texttrademark $\,$ to get 
\begin{align}
 &  \frac{1 - \hltsq_L^2}{\hlts_y^L} - \frac{ 1 - \hltsq_R^2 }{\hlts_y^R}   \nonumber \\ 
 &  =  \; \pm 2 \sinh(\lambda) \left[ \frac{\left( 1 - \hltsq_L^2 \right) \left( 1 - \hltsq_R^2 \right)}{\hlts_y^L \hlts_y^R}  + 4 \cosh^2(\lambda) \right]^{\frac{1}{2}} \!\!\!\!\!,   \label{eq:cond_kappa}
\end{align}
where $\lambda = \ln\kappa$. This reduces to \eq{eq:cond_w0} for $\kappa = 1$, since the RHS vanishes. We can also explicitly note the splitting of each solution to \eq{eq:cond_w0} into two as $\kappa$ is tuned away from 1. 

Finally, we remark that \eq{eq:cond_w2} is invariant under the substitution
\beq 
  T_L \to T_R, \quad T_R \to T_L, \quad \kappa \to \frac{1}{\kappa}. 
\eeq 
Since $\kappa$ is the coupling across the interface, this can be thought of as a ``duality'' between the strong and the weak coupling regime, albeit in a noninteracting context. Note that this duality also switches the eigenvalues, so that the physical states are not invariant under it. Instead, it is the curves defined by the matching conditions that are invariant, and the physical states in the two cases correspond to two branches of the same curve.

\begin{figure}[b]
	\centering
	\includegraphics[ width=0.7\columnwidth]{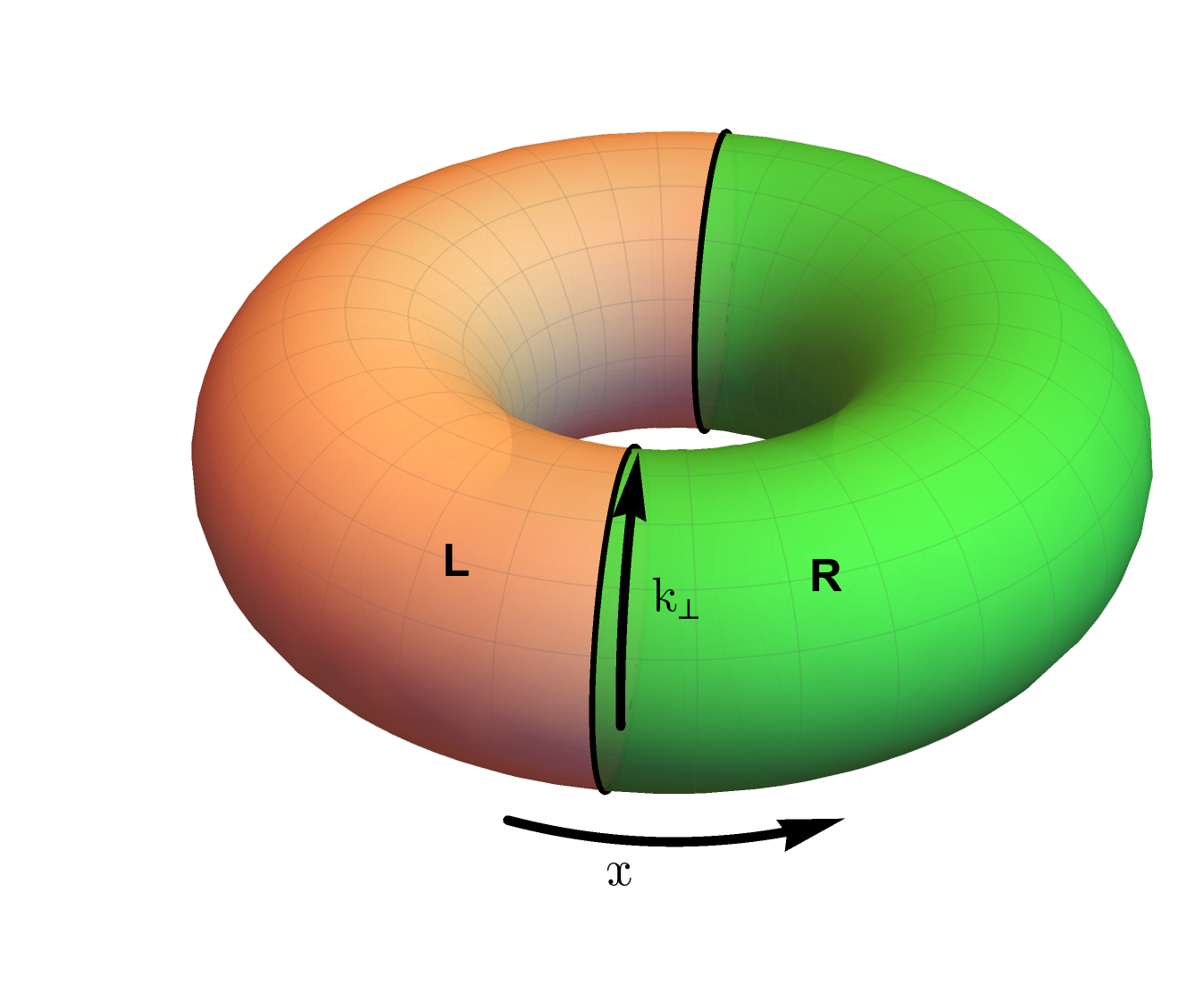}
	\caption{
	  A schematic depiction of the system used for the ED. The system is inhomogeneous along $x$ with two interfaces between the two WSMs depicted in orange and green. The transverse direction corresponds to $\vk_\perp \in \torus^2$. 
	}
	\label{fig:ed_sch}
\end{figure}
 
\subsubsection{Interface with an on-site potential} 
Finally, we consider the case of a single layer on-site potential at the interface to illustrate the computation of $S$. Explicitly, we take $W = 1$ and set
\[
  J_n = J, \quad 
  M_n = \begin{cases} 
	  M_L, & n \leq -1, \\
	  \wt{M}_0, & n = 0, \\
	  M_R, & n \geq 1,    
        \end{cases}  
\]
as shown in lower panel of \fig{fig:int_eg}. Here, $\wt{M}_0$ represents an on-site potential at layer $0$, which can model irregularities and/or other local effects at the interface. The recursion relation for $n=0$ can be rewritten as  
\beq 
  J \Psi_{1} + J^\dagger \Psi_{-1} = (\ve \id - \wt{M}_0) \Psi_0,
\eeq
which can be rewritten as $\Phi_1 = \wt{T}_0 \Phi_0$ following the construction of Sec.~\ref{sec:wsm}. Substituting $S = \wt{T}_0$ in \eq{eq:cond_det} results in the matching condition.

\begin{figure*}
	\centering
	\includegraphics[ width=0.16\textwidth]{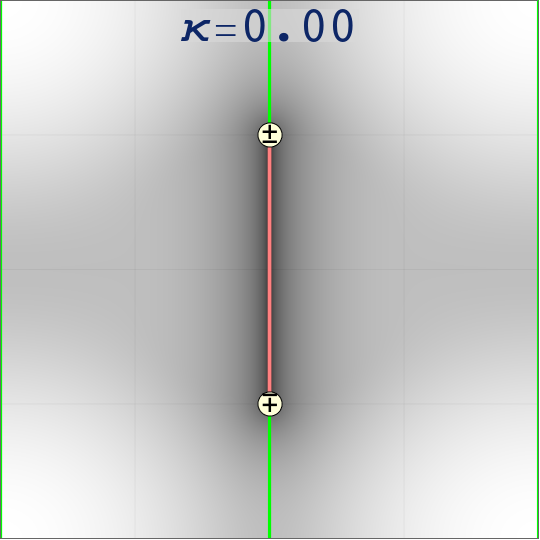}
	\includegraphics[ width=0.16\textwidth]{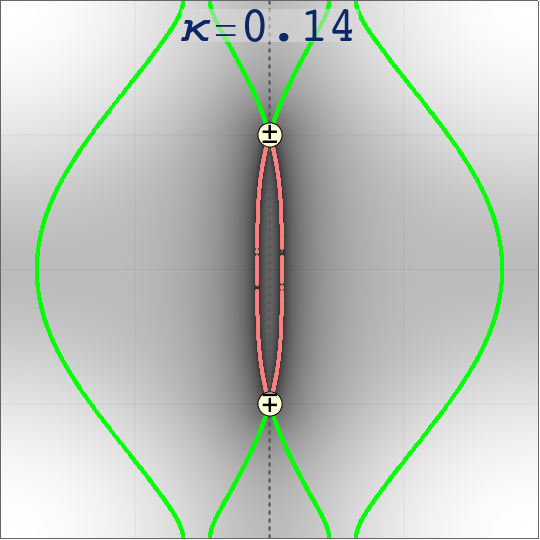}
	\includegraphics[ width=0.16\textwidth]{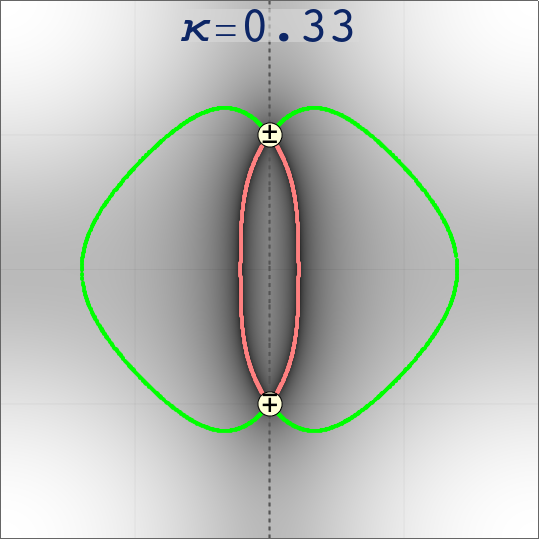}
	\includegraphics[ width=0.16\textwidth]{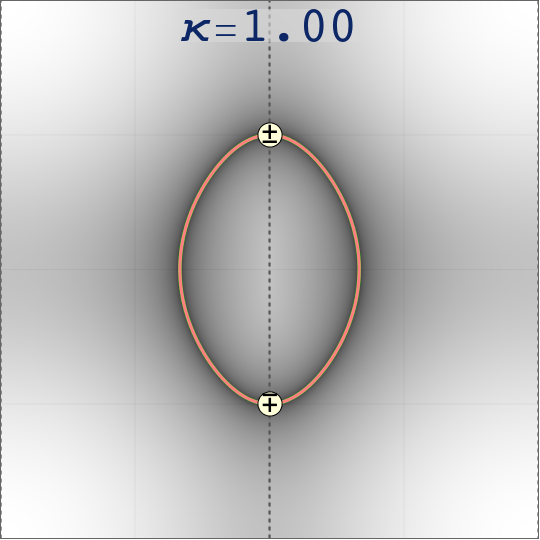}
	\includegraphics[ width=0.16\textwidth]{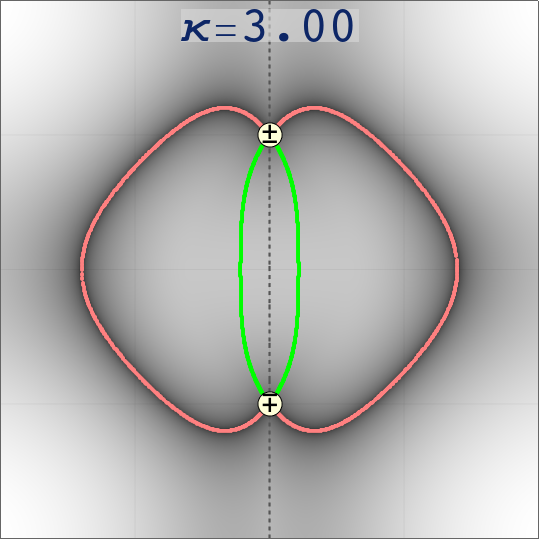}
	\includegraphics[ width=0.16\textwidth]{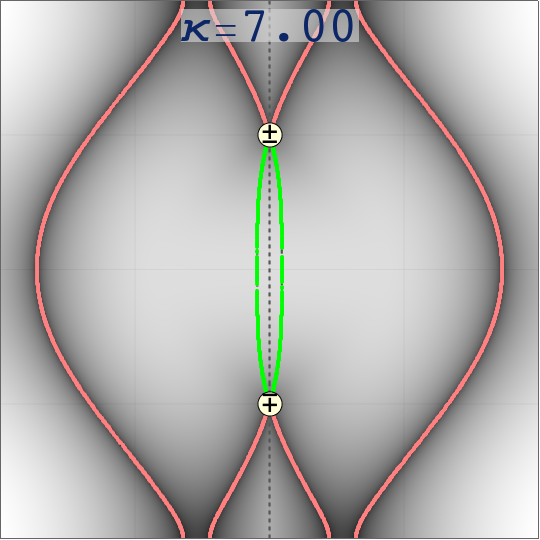}
	\caption{
	  The analytically computed locii of zero modes on the interface BZ, superimposed on the ED result for the 2-node WSM (\sect{sec:eg_2}) for $\kappa = 0, 1/7, 1/3, 1, 3, 7$ (left to right). The background depicts the lowest conduction band on the surface BZ obtained from ED, with darker shades denoting lower energies. The curves in red and green depict the decaying and growing branches of the curve obtained from the matching condition, while the dashed gray line denotes the remaining solutions. Note that the curves for $\kappa$ and $1/\kappa$ are identical, but the growing and decaying labels have exchanged their roles.
	}
	\label{fig:2node_1}
\end{figure*}


\section{Examples}     \label{sec:eg}
In this section, we derive the interface zero modes for specific models for WSMs and the interface described in \sect{sec:int_eg_2}. We compare the analytical results with the interface modes obtained by numerically diagonalizing the Hamiltonian (as a function of the transverse momentum) on a finite lattice in real space. Since a finite system with a single interface necessarily has two open surfaces, it would lead to additional modes in the spectrum obtained from ED. We instead put the system on a 3-torus, with two interfaces between the topological phases, as depicted schematically in \fig{fig:ed_sch}. 

We shall restrict ourselves to the case of tunable couplings, in which case the modes localized at the two interfaces have identical zero modes. To see this, note that the RHS of \eq{eq:cond_kappa} is invariant $L \leftrightarrow R$, while the LHS flips sign. Consequently, one switches between the two branches, switching the decay conditions, but that is precisely the correct decay conditions for the other interface.

\subsection{2-node model} \label{sec:eg_2}
We consider the simplest possible WSM, \viz, one with a single pair of Weyl nodes, described by the Bloch Hamiltonian of \eq{eq:hlt_orig}, with 
\beq 
  \bet = \left( \param \sin k_y, 1 - \cos k_y - \cos k_z \right),
\eeq 
where $\param \in \real, \, \nu \neq 0$ is a parameter. The two Weyl nodes are located at $\vk = (0,0,\pm  \pi/2)$, whose chiralities are $\pm \sgn{\param}$. The Fermi arcs are given by 
\beq 
  \sin k_y = 0, \quad 1 - \cos k_y - \cos k_z < 0,
\eeq 
which reduces to $k_y = 0,  \abs{k_z} < \pi/2$. 

Switching the sign of $\param$, the only tunable parameter in this model, switches the chiralities of the Weyl nodes and thus the orientation of the Fermi arcs. Consider then an interface between WSMs with $\param_R = -\param_L = \param$, so that the positive and negative chirality nodes swap places across the interface. To evaluate the matching condition explicitly, note that \mbox{$\hltsq_L = \hltsq_R \equiv \hltsq$}. The equation \mbox{$\hlts_y^L = \hlts_y^R$} is solved for $k_y = 0, \pi$, while the condition of \eq{eq:cond_w0} reduces to 
\beq 
  \frac{1 - \hltsq^2}{ -\param \sin k_y} = \frac{1 - \hltsq^2}{ \param \sin k_y}  \implies \hltsq^2 = 1, 
\eeq 
or more explicitly, 
\beq 
  \nu^2 \sin^2 k_y + \left( 2 + \cos k_0 - \cos k_y - \cos k_z \right)^2 = 1.   \label{eq:arc_2node}
\eeq 
Tuning away from $\kappa=1$, the matching condition from \eq{eq:cond_kappa} can be simplified to get
\beq
  \hltsq^2 = 1 \pm 2 \param \sin k_y \, \sinh \lambda,
\eeq 
where $\lambda = \ln \kappa$. In \fig{fig:2node_1}, we plot a sequence of these curve as a function of $\kappa$ alongwith the results of ED. The $\kappa \leftrightarrow 1/\kappa$ duality implies that the analytic curves obtained from the matching condition are identical for the two cases. The physical Fermi arcs, however, switch branches under  this duality 

For the decoupled WSMs ($\kappa = 0$), the Fermi arcs coincide and run along the same direction. From the intuitive arguments of \sect{sec:int}, we then expect a ``level repulsion''. Indeed, for $\kappa \leq 1$, we get a pair of curves touching at the Weyl nodes. For $\kappa > 1$, these curves expand further away from the Weyl nodes, until they hit the edges and eventually flip around the BZ around $\kappa \approx 6.85$, with additional left over curves running along the noncontractible loops of the interface BZ. In the next example, we see a more dramatic version of this detachment of lines of zero modes from the Weyl nodes. 



\begin{figure*}
	\centering
	\includegraphics[ width=0.16\textwidth]{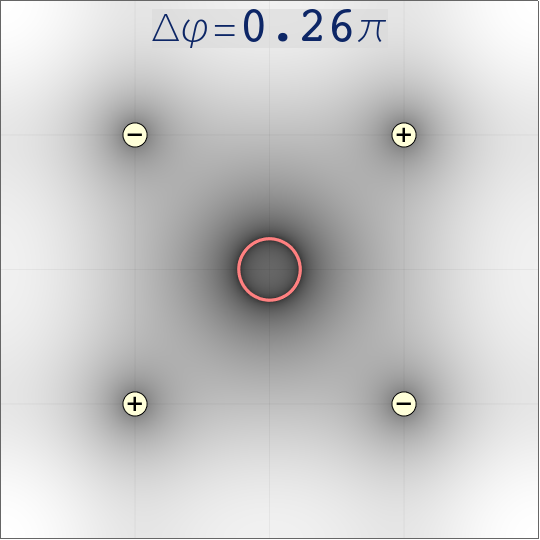}
	\includegraphics[ width=0.16\textwidth]{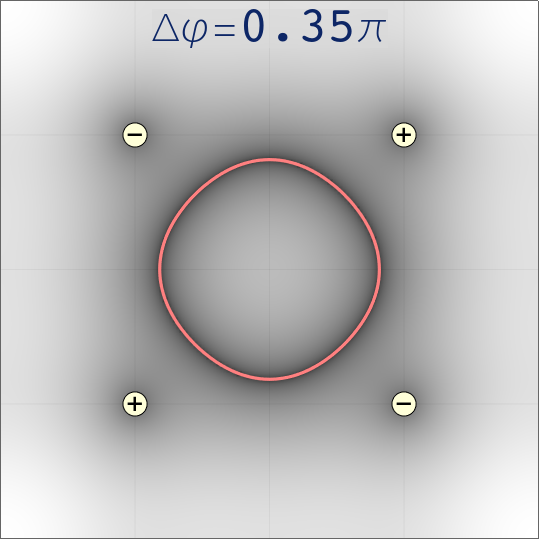}
	\includegraphics[ width=0.16\textwidth]{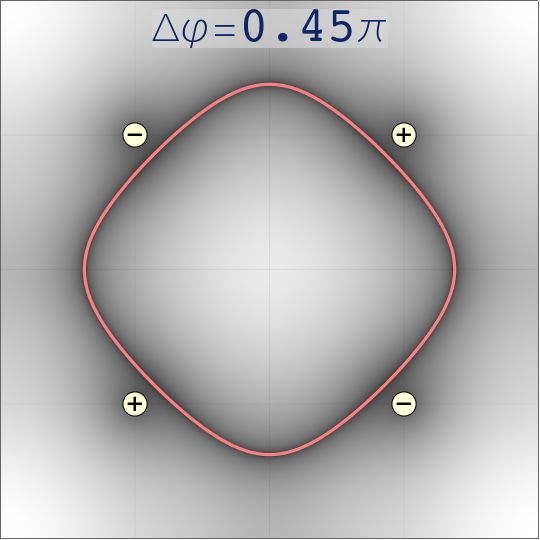}
	\includegraphics[ width=0.16\textwidth]{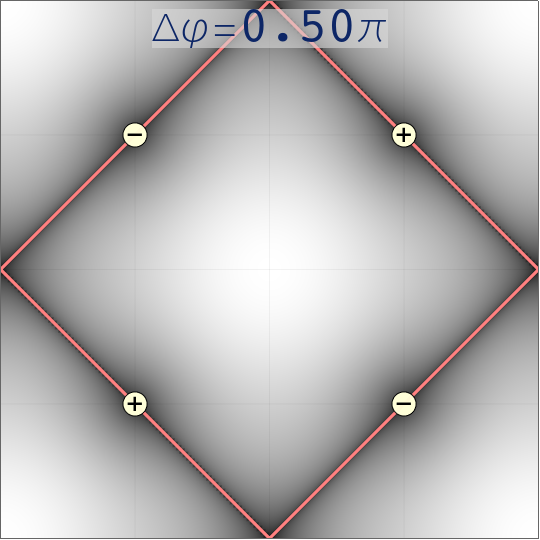}
	\includegraphics[ width=0.16\textwidth]{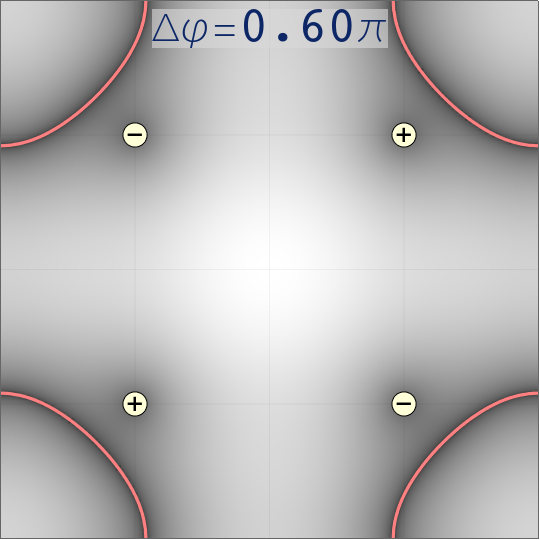}
	\includegraphics[ width=0.16\textwidth]{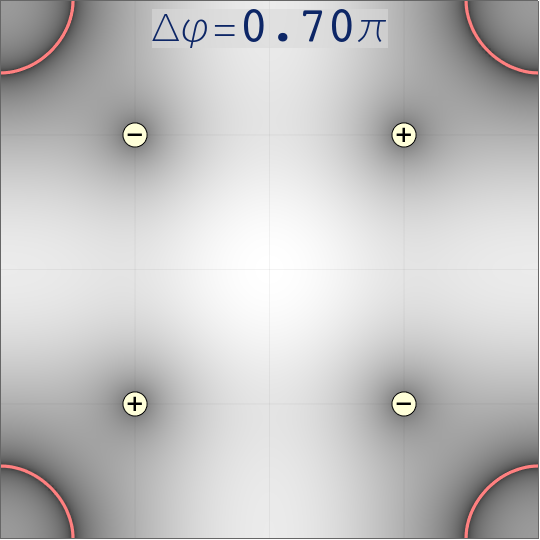}
	\ \\ \ \\ 
	\includegraphics[ width=0.16\textwidth]{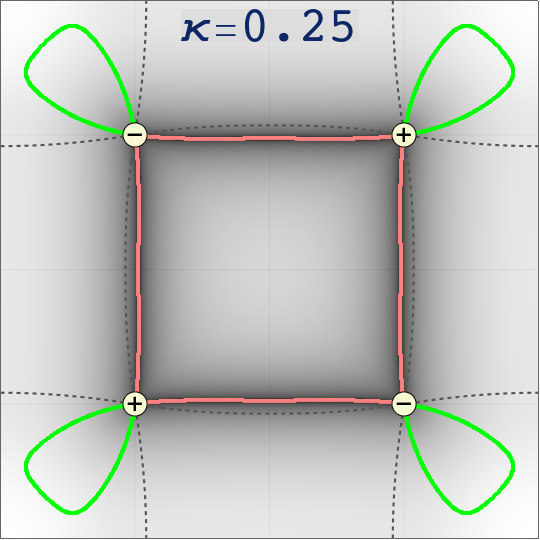}
	\includegraphics[ width=0.16\textwidth]{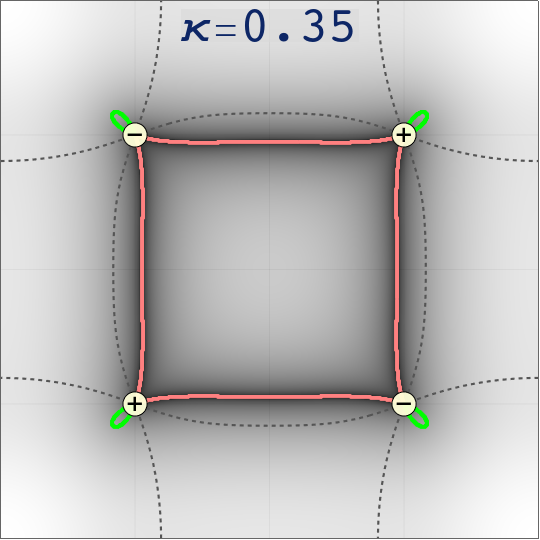}
	\includegraphics[ width=0.16\textwidth]{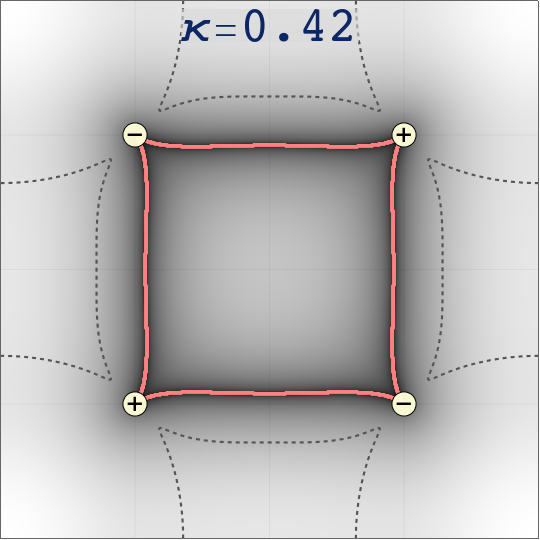}
	\includegraphics[ width=0.16\textwidth]{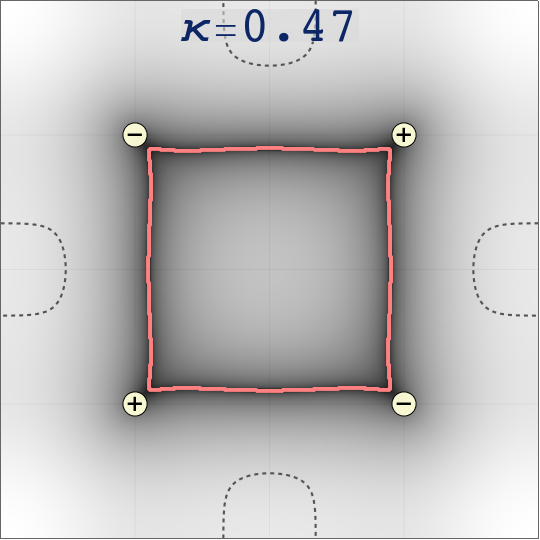}
	\includegraphics[ width=0.16\textwidth]{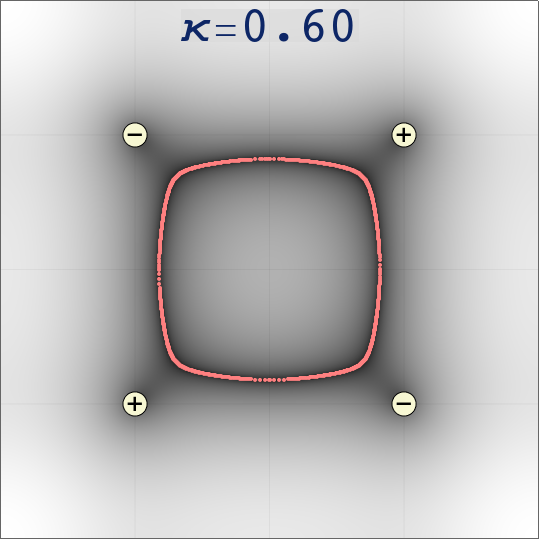}
	\includegraphics[ width=0.16\textwidth]{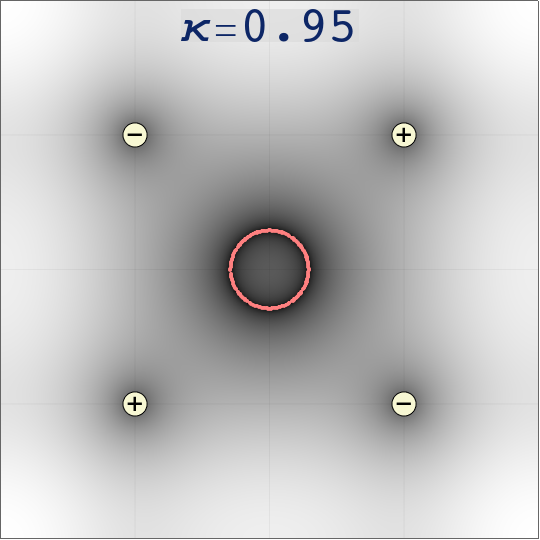}
	\caption{
	  The analytically computed locii of zero modes on the interface BZ, superimposed on the ED result for the 4-node WSM (\sect{sec:eg_4}) for a fixed $\varphi_0 = \frac{5\pi}{4}$, with (top panel) fixed $\kappa = 1$ and varying $\Delta\varphi = 0.26\pi, 0.35\pi, 0.45\pi, 0.50\pi, 0.60\pi, 0.70\pi$ from left to right, and (bottom panel) fixed $\Delta\varphi = \frac{\pi}{4}$ for $\kappa = 0.25,0.35,0.42,0.47,0.60,0.95$ from left to right. The background depicts the lowest conduction band on the surface BZ obtained from ED, with darker shades denoting lower energies. The curves in red and green depict the decaying and growing branches of the curve obtained from the matching condition, while the dashed gray line denotes the remaining solutions. For the variation of $\Delta\varphi$ as well as $\kappa$, as one tunes these parameters beyond the range explored in the plots, the interface modes gap out completely.
	}
	\label{fig:4node}
\end{figure*}

\subsection{4-node model: tunable Fermi arcs}    \label{sec:eg_4}
We next consider the family of tight-binding models proposed in Ref.~\onlinecite{vd-str_wsm}, which can be written as 
\beq 
  \bet(\varphi, \vk_\perp) = \rot(\varphi) \refvec(\vk_\perp),
\eeq
where $\varphi \in [0, 2\pi)$, $\rot(\varphi) \in \SO(2)$ is a rotation matrix and $\refvec(\vk_\perp) = (\refsym_y, \refsym_z)^T$ a reference function. The Weyl nodes are given by the zeros of $\refvec(\vk_\perp)$ independent of the parameter $\varphi$, while the Fermi arcs, defined by $\hlts_y = 0$, manifestly depend on $\varphi$. 

Consider an interface across which we change $\varphi$, so that $\bet^A = \bet(\varphi_A)$. We begin by noting that 
\beq
  \hltsq_A^2 = 1 + \abs{\bet}^2 + 2\hlts_z^A = 1 + \abs{\refvec}^2 + 2\hlts_z^A,
\eeq
since $\rot$ is a rotation matrix. The matching condition of \eq{eq:cond_w0} becomes 
\beq
  \abs{\refvec}^2 \left( \hlts_y^R - \hlts_y^L \right) = 2 \left( \hlts_z^R \hlts_y^L - \hlts_z^L \hlts_y^R \right).
\eeq
Defining 
\[
   \varphi_0 = \frac{\varphi_R + \varphi_L}{2}, \quad 
   \Delta \varphi = \frac{\varphi_R - \varphi_L}{2},
\]
we get 
\begin{align}
   \hlts_y^L - \hlts_y^R = & \; 2 \sin(\Delta\varphi) \left( \refsym_y \sin\varphi_0  + \refsym_z \cos\varphi_0 \right),	 \nonumber \\ 
   \hlts_z^R \hlts_y^L - \hlts_z^L \hlts_y^R = & \; 2 \sin(\Delta\varphi) \cos(\Delta\varphi) \abs{\refvec}^2. 
\end{align}
For an interface $\varphi_R \neq \varphi_L (\text{mod } 2\pi)$, the condition for interface modes becomes 
\beq
  \refsym_y \sin\varphi_0  + \refsym_z \cos\varphi_0 + 2 \cos(\Delta\varphi) = 0.    \label{eq:cond_4node}
\eeq
We could cancel off the prefactor $2 \sin(\Delta\varphi) \abs{\refvec}^2$, since the zeros of $\abs{\refvec}^2$ are the Weyl nodes and the zeros of $\sin(\Delta\varphi)$ satisfy $\varphi_R - \varphi_L = 2n\pi$.

In Ref.~\onlinecite{vd-str_wsm}, a particular choice of $\refvec(\vk_\perp)$ was considered:
\beq 
  \label{eq:4nodewsm}
  \vecenv{\hlts_y}{\hlts_z} = \left( 
  \begin{array}{cc} \cos\varphi & \;\; -\sin\varphi \\ \sin\varphi & \;\; \;\;\; \cos\varphi  \end{array}  
  \right) \vecenv{\cos k_y}{\cos k_z}.    
\eeq 
This model exhibits four Weyl nodes located at momenta $\vk = \left( 0, \pm \frac{\pi}{2}, \pm \frac{\pi}{2} \right)$ with chiralities as shown in \fig{fig:wsm_engg_sch}. The nodal configuration is independent of $\varphi$, while the Fermi arcs are given by 
\beq 
  0 = \hlts_y = \cos\varphi \cos k_y - \sin\varphi \cos k_z,
\eeq
which manifestly depends on $\varphi$. One can thus tune the Fermi arc connectivity using a \emph{bulk parameter}, without changing the position and chiralities of the nodes. 

We can finally solve the conundrum stated at the end of \sect{sec:int_qual}. For the coupling $\kappa = 1$, the interface modes are given by \eq{eq:cond_4node}. This has a nontrivial solution for $\vk_\perp$ iff 
\[
  \cos(\Delta\varphi) \leq \frac{1}{\sqrt{2}} \implies \abs{\varphi_R - \varphi_L} \in \left( \frac{\pi}{2}, \frac{3\pi}{2} \right),
\]
in which case one gets a closed curve of zero modes on the surface BZ. Furthermore, note that except for $\Delta\varphi = \frac{\pi}{2}$, these curves do not intersect the projections of the Weyl nodes on the interface BZ, so that they are essentially 2D Fermi surfaces localized deep inside the bulk of a 3D WSM. We plot some of these curves juxtaposed on the corresponding ED results in the top panel of \fig{fig:4node}. 

This result is unexpected and particularly intriguing, since the WSMs on the two sides of the interface have identical nodal configurations and are therefore identical from a linearized theory perspective. 
We note that although there are modes localized at the interface, they are not topologically protected; however, they are \emph{stable} against ``small'' perturbations. More precisely, they can be gapped out by the addition of an on-site term at the interface, but only by shrinking them to a point. This is analogous to the case of Weyl nodes, which are stable against small perturbations but can be gapped out by a collision with another Weyl node of opposite chirality. 

We can also vary the coupling $\kappa$ for a fixed $\Delta\varphi$ to obtain a set of interesting interface nodal configurations. In the bottom panel of \fig{fig:4node}, we plot a few such cases for a fixed $\Delta\varphi = \pi/4$ and a sequence of $0 < \kappa < 1$. The curves for $\kappa > 1$ are identical to those for $1/\kappa$, and the corresponding physical states can be obtained simply by swapping the decaying and the growing branches, as in the 2-node case.




\section{Discussion}     \label{sec:conc}
This article presents a systematic analytic study of localized modes at the interface between two noninteracting topological phases described by Bloch Hamiltonians. A closed form algebraic condition for the existence of such localized modes is derived using the transfer matrix approach. The results are used to study zero modes localized at the interface between WSMs, and compared with the zero modes obtained from a numerical diagonalization of the real space Hamiltonian.

It is worth emphasizing that the results obtained in \sect{sec:int_gen} are rather general, and can be used to study interface modes between any pairs of noninteracting topological phases. This approach, being a real space method, could also be generalized to analytically study other real space structures that might support localized modes, notable examples being defects and dislocations. The essential task is the computation of the scattering matrix $S$ introduced in \sect{sec:int_gen}, which, generically being a single particle scattering problem, should be analytically tractable. 

A particular advantage of the transfer matrix method is a complete analytic knowledge of the wavefunctions associated with the bulk as well as the boundary states. Indeed, the study of real space wavefunctions can open interesting avenues for the realization of interesting topological phases in various lattice models\cite{flore_surf_state}. Furthermore, the algebraic nature of various conditions obtained in this picture can lead to alternative perspectives on the topological invariants associated with the well-known topological phases\cite{hatsugai_cbs, vd-vc_tm}. 

The geometric information associated with the Fermi arcs has already been a topic of much discussion in the literature. The example discussed in \sect{sec:eg_4} provides yet another manifestation of this aspect, where one gets stable interface modes even when the two sides have identical nodal configuration, and are hence considered ``topologically identical''. The stable 2D Fermi surface of essentially 2D zero modes thus realized on the interface BZ, alongside the inherently 3D zero modes at the Weyl nodes opens up further interesting possibilities, e.g, the superconducting instabilities of this Fermi surface. 

The interface modes, unlike the surface modes, lie deep within the 3D bulk and are generally inaccessible to experimental techniques like ARPES. However, the Fermi arcs contribute to the transport, and are needed to \emph{complete the circuit} for the cyclotron orbits in presence of a magnetic field. Signatures of this effect are visible in the quantum oscillations of the density of states\cite{potter_quant_osc}, which might be generalizable to the case of the interface modes.

The study of states obtained at the interface between various topological phases has already led to the realization of many exotic possibilities. It can reasonably be hoped that an analytical approach to this problem will lead to a further understanding of these possibilities, as well as of the topological phases themselves.

\acknowledgments
I was funded by the Deutsche Forschungsgemeinschaft (DFG) with the CRC network TR 183 (Project B03). 
I would like to thank Victor Chua, Ciar\'an Hickey and Maria Hermanns for useful discussions.

\appendix 

\section{Condition for arbitrary rank}   \label{app:cond}
Let $T_L, T_R, S \in \SL(N, \cmplx)$ such that the matrices $T_L$ and $S^{-1} T_R S$ have a shared eigenvector $\Phi_0$. Define the family of matrices
\beq 
  C_{pq} =  \left[ T_R^p, \left(S T_L S^{-1}\right)^q \right], \quad p,q \in \intg. 
\eeq 
A necessary condition for that to be true is that $\forall p, q$, 
\beq 
  C_{pq} \Phi_0 = 0  \iff \ker \left(  C_{pq} \right) \neq \{ \nullv \}. 
\eeq 
The conditions for various $p,q$ can be combined to obtain a \emph{sufficient} condition for the existence of a shared eigenvector\cite{shemesh_common_eigs} as
\beq 
  \bigcap_{p,q \in \intg} \ker \left(  C_{pq}  \right) \neq \{\nullv \}. 
\eeq 
This condition can be simplified by noting that each square matrix $M$ must satisfy its own characteristic equation $P(M) = 0$, where $P$ is a polynomial of order $n = \dim{M}$, so that $M^n$ can be written as a linear superposition of lower powers of $M$. Thus, we need $p,q$ to only run up to $N-1$. Furthermore, $p,q = 0$ is trivial, since for any matrix, $M^0 = \id$, so that the commutator vanishes, whose kernel is $\cmplx^N$. We are left with
\beq 
  \bigcap_{p,q=1}^{N-1} \ker \left(  C_{pq}  \right) \neq \{\nullv \}.    \label{eq:cond_gen}
\eeq 
Computationally, it is more convenient to rewrite this condition as 
\beq 
  \det C = 0, \quad C \equiv \sum_{p,q=1}^{N-1} C_{pq}^\dg C_{pq}^\pdg,
\eeq 
where $C$ is positive semidefinite by definition, and a zero eigenvalue of $C$ implies a zero eigenvalue for all $C_{pq}$'s. Finally, for $N=2$, we can see that $C = C_{11}^\dagger C_{11}$, so that
\beq 
  \det C = 0 \implies \det \left[ T_R, S T_L S^{-1} \right] = 0,
\eeq 
which is precisely what we obtained in Sec \ref{sec:int}.

\section{Computations for the trivial interface}     \label{app:calc}
We seek to explicitly compute $\det [T_L, T_R]$. From the definition of the transfer matrix of \eq{eq:tmat}, we may write 
\beq 
  T_A(\ve, \vk_\perp) = \frac{1}{1 + \hlts_z^A}\left( 
  \begin{array}{cc} 
    \ve^2 - \left( \hltsq^A\right)^2 & \quad  -\ve + \hlts_y^A \\ 
    \ve + \hlts_y^A & \quad -1    
  \end{array} \right),
\eeq 
where 
\[
  \hltsq_A^2 = \left( \hlts_y^A \right)^2 +  \left( 1 + \hlts_z^A \right)^2.
\]
In order to perform this computation explicitly, we separate $T_A$ into parts that depend on $A$ and parts that are independent of $A$, as 
\beq 
  T_A = \frac{1}{1 + \hlts_z^A} \left( t_0 + t_A \right),
\eeq 
where we can expand $t_0$ and $t_A$ in terms of the Pauli matrices as 
\begin{align} 
  t_0 = & \; \left( 
  \begin{array}{cc} 
    \ve^2 & \quad  -\ve \\ 
    \ve & \quad -1 
  \end{array} \right) \nonumber \\ 
  = & \; \frac{\ve^2-1}{2} \id - i\ve \sigma^y + \frac{\ve^2+1}{2} \sigma^z, \nonumber \\ 
  t_A = & \; 
  \left( 
  \begin{array}{cc} 
    -\hltsq_A^2 & \quad  \hlts_y^A \\ 
    \hlts_y^A & \quad 0 
  \end{array} \right) \nonumber \\ 
  = & \; -\frac{1}{2} \hltsq_A^2 \id + \hlts_y^A \sigma^x - \frac{1}{2} \hltsq_A^2 \sigma^z.
\end{align}
The commutators can be conveniently computed using the commutation relations of the Pauli matrices, $[ \sigma^i, \sigma^j ] = 2 i \epsilon^{ijk} \sigma^k$. We shall need
\begin{align}
 [t_0, t_A] = & \;  \left[  - i\ve \sigma^y + \frac{\ve^2+1}{2} \sigma^z, \, 
      \hlts_y^A  \sigma^x - \frac{1}{2} \hltsq_A^2 \sigma^z \right] \nonumber \\ 
 = & \; - \ve \hltsq_A^2 \sigma^x + \hlts_y^A (\ve^2 + 1) i \sigma^y - 2 \ve \hlts_y^A \sigma^z, 
\end{align}
and 
\begin{align}
 [t_L, t_R] = & \; \left[ \hlts_y^L  \sigma^x - \frac{1}{2} \hltsq_L^2 \sigma^z, \, 
      \hlts_y^R  \sigma^x - \frac{1}{2} \hltsq_R^2 \sigma^z \right] \nonumber \\ 
 = & \; \left( \hlts_y^L \hltsq_R^2  - \hlts_y^R \hltsq_L^2 \right) i \sigma^y. 
\end{align}
The condition for the interface modes becomes
\begin{align}
  0 = & \; \det\left( [t_0 + t_L, t_0 + t_R] \right) \nonumber \\ 
  = & \;  \det\left(  [t_0, t_R - t_L] + [t_L, t_R] \right)  
\end{align}
We can compute this as 
\begin{align} 
  0 = & \; \det\left(  a \ve \, \sigma^x + i (b \ve^2 + c) \, \sigma^y + d \ve \, \sigma^z \right)  \nonumber \\ 
  = & \; a^2 \ve^2 - (b \ve^2 + c)^2 + d^2 \ve^2,    \label{eq:int_cond2}
\end{align}
where we have defined the real functions $a,b,c,d$ as
\begin{align}
 a = & \; \hltsq_L^2 - \hltsq_R^2, \nonumber \\  
 b = & \; \hlts_y^R - \hlts_y^L,   \nonumber \\ 
 c = & \; \hlts_y^R \left\{ 1 - \hltsq_L^2  \right\} - \hlts_y^L \left\{ 1 - \hltsq_R^2  \right\},  \nonumber \\ 
 d = & \; 2 \left( \hlts_y^R - \hlts_y^L \right).   
\end{align}
We may solve \eq{eq:int_cond2} for $\ve$ to get 
\begin{align*}
 \ve = \frac{1}{2b} \left[ \zeta_1\sqrt{a^2 + d^2} + \zeta_2 \sqrt{a^2 + d^2 - 4bc}   \right],
\end{align*}
where $\zeta_1, \zeta_2 = \pm 1$. These four solutions give the spectrum of the interface modes for the 2-band model in question.

\section{Boundary vs decay conditions}  \label{app:unphys}
Consider the case of a lattice with open boundary conditions along $x \in [0,N]$ and periodic boundary conditions along the remaining (transverse) direction. Let $T(\ve, \vk_\perp) \in SL(2,\cmplx)$ be the transfer matrix for translations along $x$, so that $\Phi_n = T^n \Phi_0$. The open boundary conditions for $\Phi_n$ demand that 
\beq 
  \Phi_0 \propto \vecenv{1}{0} \equiv \vartheta_L , \quad   
  \Phi_N \propto \vecenv{0}{1} \equiv \vartheta_R.   \label{eq:bound_cond}
\eeq 
Given any vector $\vartheta \in \cmplx^2$, we can use it to define a ``state'' with $\Phi_n = T^n \vartheta$. For this state to be localized at the boundary, we demand that: 
\begin{enumerate}
 \item $\vartheta$ satisfies the relevant boundary conditions, i.e, $\vartheta \propto \vartheta_L$ for a left edge state and  $\vartheta \propto \vartheta_R$ for a right edge state. 
 
 \item $\norm{\Phi_n} = \norm{T^n \vartheta} \to 0$ as $n\to\infty$ for a left edge state, and as $n\to -\infty$ for a right edge state.
\end{enumerate}
We shall refer to these as the \emph{boundary} and \emph{decay} conditions, respectively. 

For a $2\times 2$ transfer matrix, since the product of the eigenvalues is 1, the decay condition implies that $\vartheta$ must be an eigenvector\cite{vd-vc_tm} of $T$. Then, 
\beq 
  T \vartheta = \rho \vartheta \implies \norm{\Phi_n} = \abs{\rho}^n \norm{\Phi_0},
\eeq 
so that $\abs{\rho}$ controls the asymptotic behavior of the state for $n\to \pm \infty$. More precisely, we have a left edge state if $\rho$ lies within the unit circle, and a right edge state if it lies outside the unit circle. Putting together the boundary and decay conditions, we get the following table:
\begin{table}[ht!]
  \vspace{0.1in}
  \renewcommand{\arraystretch}{1.5}
  \setlength{\tabcolsep}{20pt}
  \begin{tabular}{|c|ll|}
    \hline 
     & $\abs{\rho}<1$ & $\abs{\rho}<1$ \\ 
     \hline
     $\vartheta \propto \vartheta_L$ & Left edge  & \emph{Unphysical} \\ 
     $\vartheta \propto \vartheta_R$ & \emph{Unphysical} & Right edge \\ 
    \hline  
  \end{tabular}
  \caption{Boundary(rows) vs decay(column) conditions.}   \label{tb:edge}
\end{table}

Choosing a single row/column of this table corresponds to a single condition on $(\ve, \vk_\perp)$. If we choose the conditions for individual of this table, we shall obtain the set of curves which depict the set of modes localized at the  boundaries for all possible choices of boundary conditions. We claim that the same is true if we choose the condition for a row. To see this, consider the first row. Then, if we get $\abs{\rho}<1$, the state is a physical state, localized at the left edge. For $\abs{\rho}>1$, the states grows as $n\to\infty$, however, it can be localized at the right edge instead, if we were to choose a new boundary condition for the right edge, \viz, $\wt{\vartheta}_R = \vartheta_L$. Physically speaking, this may sometime correspond to an unusual boundary condition, however, mathematically it is well defined. 

Thus, to expose the complete topological information encoded in the edge spectra, we must choose all four possibilities listed in Table~\ref{tb:edge}. The easiest way to implement this is to simply consider the conditions for the two rows, which, using \eq{eq:bound_cond}, reduce to an entry of the transfer matrix being zero. These level sets can be interpreted as algebraic curves\cite{hatsugai_cbs, vd-vc_tm} in (a complexified) $\ve$, which can be used to define certain winding numbers, which are topological invariants associated with the boundaries. 

Turning to the case of an interface, the boundary conditions are now replaced by a \emph{matching} condition, which we derived explicitly as 
\beq 
  \det[T_L, S^{-1} T_R S] = 0. 
\eeq 
The analysis for a finite system readily generalizes, except that we shall now need to tune the gluing matrix instead of the boundary vectors. More precisely, consider a state defined by $\vartheta \in \cmplx^2$ with the \emph{wrong} decay condition, i.e, 
\begin{align}
 & T_R S \vartheta = \rho_R S \vartheta,  & \!\!\!\!\!\!\!\!\!\!\!\!\!\!\!\! \abs{\rho_R} > 1,  &  \quad \nonumber \\ 
 & T_L \vartheta = \rho_L \vartheta,  & \!\!\!\!\!\!\!\!\!\!\!\!\!\!\!\! \abs{\rho_L} < 1.  &  \quad
\end{align}
This instead represents a physical localized mode at the interface with the right and left sides swapped, and the gluing matrix $S$ replaced by $S^{-1}$. More explicitly, we observe that under defining
\[
 \wt{T}_R = T_L, \quad 
 \wt{T}_L = T_R, \quad 
 \wt{S} = S^{-1},  
\]
we note that $\wt{T}_R$, $\wt{T}_L$ and $\wt{S}$ still satisfy the matching conditions, while the corresponding eigenvalues $\wt{\rho}_R$ and $\wt{\rho}_L$ now satisfy the \emph{correct} decay conditions. 



\bibliography{wsm_int}

\end{document}